# Optoelectronic Study of (MA)$_2$NaBiX$_6$ (MA= methylammonium; X=Cl,Br, I) Hybrid Double Perovskites by Ab initio Simulation

**A. Johnson, T. J. Ikyumbur and F. Gbaorun**
Department of Physics, Benue State University, Makurdi, Nigeria.
*Corresponding Author's Email:* ajohnson17@bsum.edu.ng

**Abstract**
High prospects for different optoelectronic applications have drawn much effort towards hybrid double perovskites for the commercialization of environmentally friendly Pb-free non-toxic perovskites. Herein, the optoelectronic study of (MA)$_2$NaBiX$_6$ (MA= methylammonium; X=Cl,Br, I) hybrid double perovskites by ab initio simulation was investigated. (MA)$_2$NaBiBr$_6$ and (MA)$_2$NaBiCl$_6$ have indirect X-M bandgaps while (MA)$_2$NaBiI$_6$ have indirect X- Γ bandgap of 2.98 eV, 3.64 eV and 2.32 eV respectively. The results also show that the iodide-based compound, (MA)$_2$NaBiI$_6$ exhibits superior optoelectronic properties compared to the bromide and chloride-based compounds, (MA)$_2$NaBiBr$_6$ and (MA)$_2$NaBiCl$_6$. These findings indicate that the (MA)$_2$NaBiI$_6$ hybrid double perovskite is a potential material for optoelectronic applications owing to its high absorption coefficient (on the order of 10$^6$cm$^{-1}$), dielectric constant (approx.3.34), and refractive index (2.50) in addition to its high formation energy, which shows its stability.

## 1 Introduction

Perovskite materials are gaining popularity in a number of applications such as high temperature superconductivity, ferromagnetism, ferroelectricity, piezoelectricity, photocatalysis and photovoltaic (Filip, Filip, & Volonakis, 2018; Liu et al., 2022; Volonakis et al., 2016). These novel materials have drawn great attention as a result of the discovery of different families of lead halide perovskite materials used as light absorbers in solar cells showed an increase in the power conversion efficiency (PCE) from 3.8% to 22.7% in less than a decade as compared to their predecessors, the dye-sensitized solar cells (DSSCs) which recorded a maximum efficiency of 11.9% over a period of two decades of research (Meyer, Mutukwa, Zingwe, & Taziwa, 2018). Recently, a PCE value of 24.86% was obtained for lead-free perovskite (Hossain, Hasan, Rahman, & Munaim Hossain, 2020; Lin, 2020), and looking at the processing mechanisms for the production of these materials being simple aqueous chemistry which is far cheaper than growing crystals by melting or vapor deposition, it is obvious that these solution-processed perovskite materials are the leading candidates for the next generation of solar cells (Filip et al., 2018; Liu et al., 2022; Nelson & Cochran, 2019; Song et al., 2015). Nonetheless, the breakthroughs recorded by these perovskite materials have not yielded the desired commercialization of these solar cell materials due to a number of challenges such as degradation owing to moisture and heat, as well as prolonged exposure to light in addition to their proneness to ion or halide vacancy migration which result in distortion in their photovoltaic usability (Filip et al., 2018; Hoke et al., 2015; Manser, Saidaminov, Bakr, & Kamat, 2016; Meloni et al., 2016), also, the toxicity of lead poses a great environmental concern (Ning & Gao, 2019), and the stability of the materials (Liu et al., 2022). Evidently, to attain commercialization of perovskite-based devices, these challenges need to be addressed (Korbel, Marques, & Botti, 2018; Palummo, Berrios, Varsano, & Giorgi, 2020; Schade et al., 2019).

Earlier studies in this direction focused on lead replacement with elements such as germanium or tin which are to be in the same group as lead on the periodic table. However, the tin-based perovskites suffered from degradation in air (M Roknuzzaman et al., 2019). Additionally, germanium-based perovskites showed poor photovoltaic performance with low light absorption, dielectric constant and optical conductivity (M Roknuzzaman et al., 2019). Lead-free perovskite materials based on divalent elements other than group-IVA exhibit poor optoelectronic properties (Zhang, Liao, & Yang, 2019). These pitfalls obviously show that the development of lead-free perovskites for optoelectronic applications remains an open challenge (Dong et al., 2023).

Recently, it has been observed that methylammonium (MA) based organic–inorganic hybrid double perovskites offer better optoelectronic properties than Cs contained inorganic perovskites in addition to depicting similar optoelectronic properties with $Cs_2BiAgI_6$, $Cs2SbAgI_6$ and $MAPbI_3$ (Varadwaj & Marques, 2020). These hybrid perovskites are also known to exhibit more negative formation energies compared with the inorganic perovskites which posits that they are more stability (Roknuzzaman, 2020). However, from these studies, the use of a noble metal Cu or Ag is evident and due to depletion of the copper ore (Rötzer & Schmidt, 2018) as well as the high cost of silver, Na is employed in this study as a substitute of the noble metal in addition to investigating different combination to find suitable replacement for Pb in double perovskites for different applications. This is because $Ag^+$ with 1.15 Å ionic radius can be replaced with $Na^+$ (1.02 Å) due to the similarity of their ionic radius (Shuai & Ma, 2016; Volonakis et al., 2016). Specifically, in order to attain commercialization of these novel materials and reduce the cost associated with noble metal-based precursors thereby boosting their usability, optoelectronic study of $(MA)_2NaBiX_6$ (MA= methylammonium; X=Cl, Br, I) hybrid double perovskites by Ab initio simulation is investigated.

## 2 Materials and Method

## 2.1 Computational Method

The first-principles density functional theory (DFT) simulations were performed using Quantum ESPRESSO (QE) package (Lab, 2021) with the ultrasoft pseudopotential including core correction. The Perdew-Berke-Ernzerhof (PBE) procedures under the local density approximation (LDA) have been chosen to evaluate the exchange and correlation effects of the valence electrons (Ropo, Kokko, & Vitos, 2008). In optimizing the crystal structure and estimating the ground state energy, the Broyden-Fletcher-Gordfarb-Shanno (BFGS) minimization technique has been used in QE. All through the calculations, the C (2s22p2), H (1s1), I (5s25p5), N (2s22p3), Na (3s1), Bi(5d106s26p3) valence electronic configurations were used. A $k$ point of 4×4×4, plane-wave cutoff energy of 30 Ry and charge density of 180 Ry were applied in obtaining the best convergence and total energy value used for all calculations associated with LDA-PBE. The cells were fully relaxed until the cell stresses were at least $10^{-6} Ry/bohr^3$ and the residual forces on each atomic site were at least $10^{-4} Ry/bohr$. In addition, neither the spin−orbit coupling effect nor HSE06 functional was considered in the electronic structure calculations in this study. This is due to error cancellation, such that by neglecting the spin−orbit interaction results in band gap overestimation, which in turn, cancels the band gap underestimation error of lower approximations such as LDA calculation (Even, 2013).

## 2.2 Formation Energy Calculation

To determine the thermal stability of these perovskites, the formation energies, $E_{form}$ for the perovskites were calculated using Equation (1):

$$E_{form} = E_{tot} - \Sigma_x E_{tot}(x) \qquad (1)$$

Where, $E_{form}$ is the formation energy of the perovskite; $E_{tot}$ is the total energy of the perovskite compound and $\Sigma_x E_{tot}(x)$ is the sum of the total energy components of the perovskite compound which were obtained from the energy calculation in the optimization of the compounds.

## 3 Results and Discussion

### 3.1 Structural properties

Table 1. Calculated lattice constants and bandgaps for hybrid double halide perovskites $(MA)_2NaBiX_6$ (X = Cl, Br, I).

| Compounds | ɑ (Å) | Bandgap (eV) (LDA-PBE) |
|---|---|---|
| $(MA)_2NaBiCl_6$ | 7.82 | 3.64 |
| $(MA)_2NaBiBr_6$ | 8.24 | 2.98 |
| $(MA)_2NaBiI_6$ | 8.85 | 2.32 |

Table 1 shows the calculated lattice parameters of the hybrid double perovskites. The values appear to be changing with the substitution of halogen atom down the group as depicted by the increasing values from the Cl-based compound to I-based compound. The variation of the optimized lattice parameters with bandgap energy values in the present study is in good agreement with (Greenman, Williams, & Kioupakis, 2019; Johnson, A, Gbaorun, F and Ikyo, 2022). The atomic size of the halogen atoms increases down the group, thereby increasing the concentration of the halide ( Kragh, 1963;Yao et al., 2021). This in turn, decreases the value of the bandgap energy down the group (M Roknuzzaman, Ostrikov, Wang, Du, & Tesfamichael, 2017; M Roknuzzaman et al., 2019).

### 3.2 Electronic properties

Results of the electronic band structures for the hybrid double perovskites as shown in Figures 1-3 reveal that these hybrid double perovskites were predicted to be of indirect band gap as the conduction band minima (CBMs) and valence band maxima (VBMs) appear at different points on high symmetric path of the BZ. These observations are similar to those of the inorganic double perovskites, $Cs_2NaBX_6$ (B = Bi, Sb; X= Cl, Br, I) (Shuai & Ma, 2016). Generally, the electronic bandgap energy ($Eg$) can be obtained from the difference between CBM and VBM in the electronic band structures as given in equation (2):

$$Eg = CBM - VBM \qquad (2)$$

The CBMs were shown to appear at the X point of the high symmetric path of the BZ. On the other hand, the VBMs of these materials were found to be at the M point with exception of $(MA)_2NaBiI_6$ with VBM at Γ. The calculated band gap values are presented in Table 1 from which it can be inferred that the smallest bang gap value was obtained for the iodide hybrid double perovskite $(Ma)_2NaBiI_6$ (2.32 eV). By comparison, this value show similarity with 2.19 eV of $Cs_2AgBiBr_6$ obtained through measurement techniques by (McClure, Ball, Windl, & Woodward, 2016). The perovskites in the present method gave band gap values of 2.32 eV, 3.64 eV and 2.98 eV for $(Ma)_2NaBiI_6$, $(MA)_2NaBiCl_6$ and $(MA)_2NaBiBr_6$ respectively. These values are comparable with that of $(MA)_2BiAgCl_6$ (2.7 eV) obtained by (Volonakis et al., 2016) within the same level of theory and that of $MAPbCl_3$ (3.0eV) by (McClure et al., 2016) with an exemption with $(MA)_2NaBiCl_6$ perovskite with 3.7eV in the present study with some deviations which could be as the result of its small value of lattice parameter as observed by (Greenman et al., 2019).

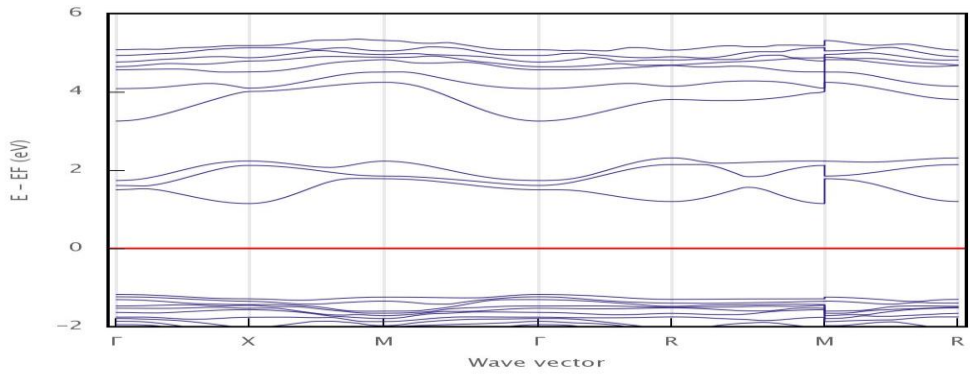

Figure 1. Electronic band structure of $(MA)_2NaBiI_6$. Bandgap energy
$E_g$= 2.32 eV with CBM at X and VBM at Γ in the LDA-PBE approximation

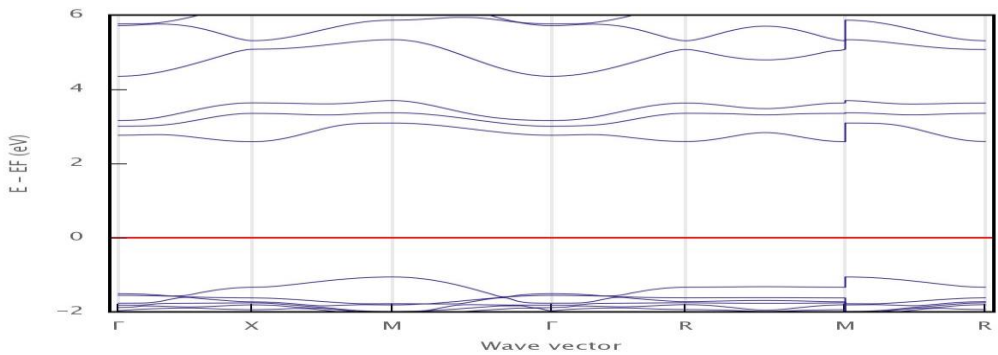

Figure 2. Electronic band structure of $(MA)_2NaBiCl_6$. Bandgap energy
$E_g$= 3.64 eV with CBM at X and VBM at M in the LDA-PBE approximation

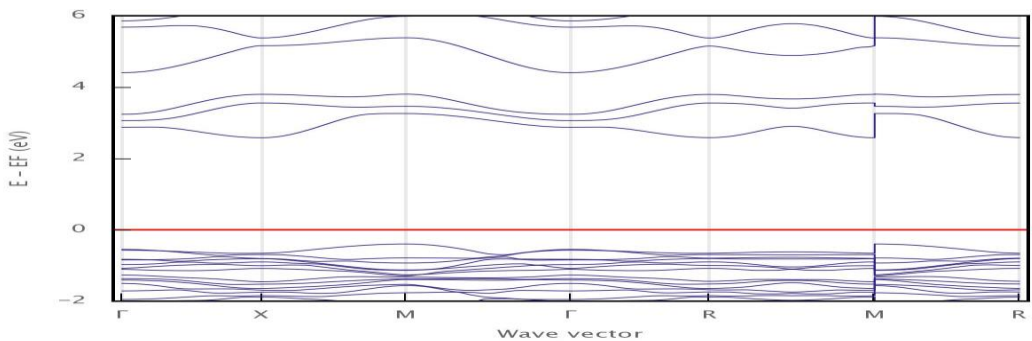

Figure 3. Electronic band structure of $(MA)_2NaBiBr_6$. Bandgap energy
$E_g$= 2.98 eV with CBM at X and VBM at M in the LDA-PBE approximation

The total and projected density of states (DOS) of the $(MA)_2NaBiI_6$ perovskite is shown in Figure 4. Here, it can be seen that the Bi-6p states and I-5p states are the major contributors to the total DOS towards VBM while the contribution to the total DOS towards CBM is by the I-5p states only. The Na-3s states contribute only partly to the total DOS towards VBM. This is similar to the Na atom in the inorganic double perovskites, $Cs_2NaBX_6$ (B=Bi, Sb; X= I, Br, Cl) investigated by (Shuai & Ma, 2016). They observed that while Na atom is significant in the formation of the double perovskite crystal, the Na *s* orbital is not involved in the composition of VBM of the perovskite material.

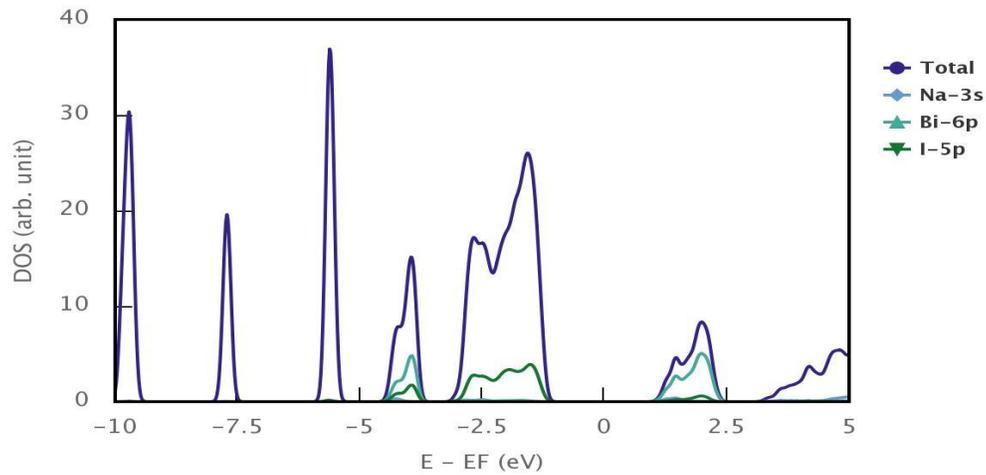

**Figure 4.** Calculated and projected density of states of $(MA)_2NaBiI_6$ perovskite shows that the contribution to the total DOS towards VBM is by the Bi-6p states and I-5p states. The contribution to the total DOS towards CBM is by the I-5p states while the Na-3s states gave apartial contribution to the total DOS towards VBM.

### 3.3 Optical properties

The calculated dielectric functions, absorption coefficients, real parts of the refractive indices of $(MA)_2NaBiX_6$ (X = Cl, Br, I) were calculated using Equations (3), (4) and (5) respectively (Shuai & Ma, 2016), and presented in Figures 5(a-c).

(a)

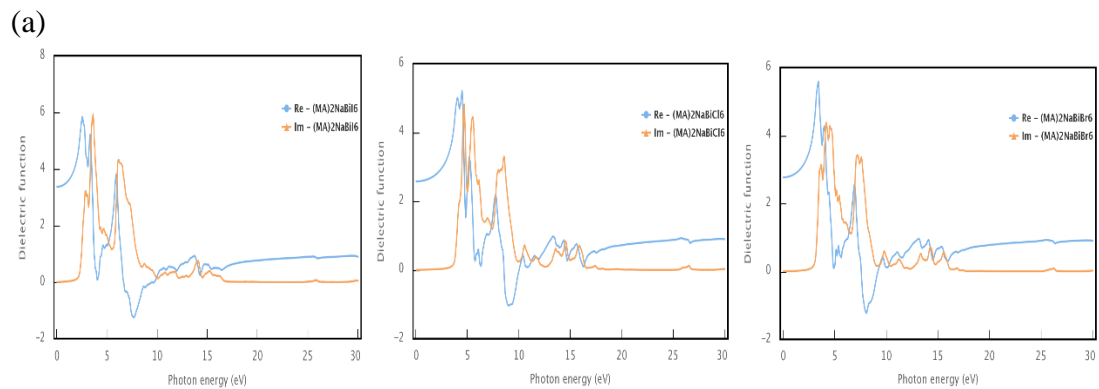

(b)

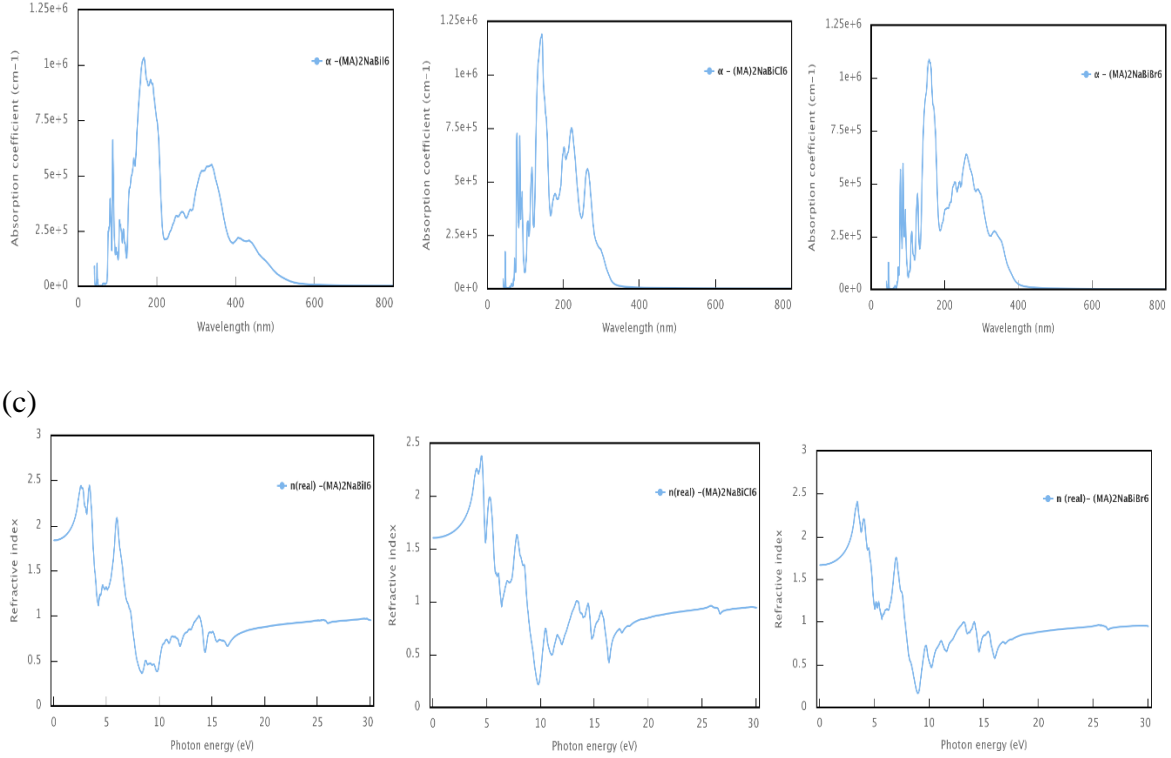

**Figure 5.** The optical properties of double perovskites (MA)$_2$NaBiX$_6$ (X = Cl, Br, I) along the incident electromagnetic radiation of energy from 0 to 30 eV. (**a**) Calculated dielectric function (**b**) Calculated absorption coefficient (**c**) Calculated refractive index.

The optical response between occupied and unoccupied orbitals are described by the complex dielectric function $\varepsilon(\omega)$ (Shuai & Ma, 2016):

$$\varepsilon(\omega) = \varepsilon_r(\omega) + i \cdot \varepsilon_i(\omega) \qquad (3)$$

where the imaginary part, $\varepsilon_i(\omega)$ reveals the empty states of the electronic structure as well as the absorptive characteristic. From this, the optical property of materials can be easily estimated from $\varepsilon(\omega)$ based on the following relations (Shuai & Ma, 2016):

$$n(\omega) = \sqrt{\frac{\sqrt{\varepsilon_r(\omega)^2 + \varepsilon_i \omega^2} + \varepsilon(\omega)}{2}} \qquad (4)$$

$$\alpha(\omega) = \frac{4\pi}{\lambda}\sqrt{\frac{-\varepsilon_r(\omega) + \sqrt{\varepsilon_r^2(\omega) + \varepsilon_i^2(\omega)}}{2}} \qquad (5)$$

Equations (4) and (5) give the refractive index and absorption coefficient respectively. Where $\lambda$ and $\omega$ represent the wavelength and frequency of the incident light respectively.

The dielectric function and refractive indices from Fig. 5 (a, c), were analyzed for the photon energy between 0 and 30 eV; while the absorption coefficients were investigated for the wavelength from 0 to 800 nm as depicted in Fig. 5(b). As can be seen in Fig. 5(a), the intensity of the imaginary part of the dielectric function of all samples varies in height from Iodine incorporated compound at 3.5 eV phonon energy to 4.0 eV and 4.5 eV for Bromine and Chlorine incorporated compounds respectively. This agrees with the trend obtained for the bandgap values for all the samples, implying that both the photon energy of the dielectric function and the bandgap of these three analyzed incorporated halogens follow same trend in value (3.5<4.0<4.5) eV for $I_6<Br_6<Cl_6$ respectively. This agrees with (Rajivgandh, Govindan Nadar Chackaravarthi et al., 2022) on the variation of bandgap energy value with dielectric constant.

Owing to the dependency of the absorptive behavior of a material on the imaginary part of the dielectric function (Shuai & Ma, 2016; Varadwaj & Marques, 2020), the imaginary parts of the dielectric function were investigated. The material $(MA)_2NaBiI_6$ showed the highest value for the imaginary dielectric function of all the investigated perovskites materials for solar radiation and suggests that it is a potential optoelectronic material.

Generally, the optoelectronic performance of a material depends on high dielectric constant or dielectric function at zero frequency. As opined by (Johnson, A, Gbaorun, F and Ikyo, 2022; Ramanathan & Khalifeh, 2021; Varadwaj & Marques, 2020) that materials with large dielectric constant can hold large amount of charge over a long period of time and as such, enhancing the optoelectronic performance. In this study, the highest value of the dielectric constant of 3.34 is obtained for $(MA)_2NaBiI_6$ material and shows a variation of about 1.09 with 4.43 obtained for the $(CsMA)NaSbI_6$ material (Johnson, A, Gbaorun, F and Ikyo, 2022). This suggests that other optical properties of the hybrid double perovskite $(MA)_2NaBiI_6$ material could be a suitable substitute to Pb-based hybrid perovskites for optoelectronic applications.

According to Varadwaj & Marques (2020), absorption coefficients that are less than $10^4 cm^{-1}$ depict weak absorption, however, as can be seen in the present study, the compounds show absorption coefficients of the order of $10^6 cm^{-1}$ with the highest absorption obtained by $(MA)_2NaBiI_6$. Nonetheless, as shown in Figure 5b, the absorption coefficient decreases with increasing wavelength and is not uniform as observed by (Jin, 2013). This is expected as revealed in Equation (5) due to the inverse dependence of $\alpha$ on the wavelength $\lambda$ as well as the fact that the probability of a photon being absorbed depends on the likelihood of interacting with an electron. In Figure 5(c), the refractive index for $(MA)_2NaBiI_6$ showed the maximum peak (2.50) over a broad spectrum (2-4 eV) which suggests that the $(MA)_2NaBiI_6$ material can be suitable in optoelectronic applications including organic-light emitting diode (OLED), liquid crystal display (LCDs), and quantum dot light emitting diode (QDLED) televisions where materials with high refractive index (>1.50) (Garner, 2019; Johnson, A, Gbaorun, F and Ikyo, 2022) are needed.

### 3.4 Thermal properties

Table 2. Calculated formation energy, H(kJ/mol) of $(MA)_2NaBiX_6$ (X = Cl, Br, I)

| Compounds | H(kJ/mol) |
| --- | --- |
| $(MA)_2NaBiCl_6$ | -8,036.3 |
| $(MA)_2NaBiBr_6$ | -7,867.9 |
| $(MA)_2NaBiI_6$ | -7,627.0 |

From equation (1), the formation energies of the perovskites compounds were calculated. As can be inferred form Table 2, the investigated perovskites posit high negative formation energies which increase down the group from Cl to I. These reveal the stability of the materials and increase with increasing negativity of the formation energy. Among all the investigated compounds, the formation energy increases in the order of $(MA)_2NaBiCl_6 <$ $MA)_2NaBiBr_6 < (MA)_2NaBiI_6$, making I-based compound the most stable (Mutter & Urban, 2020; Roknuzzaman, 2020) These results, in essence, show that the compounds can be successfully synthesized with potential for a number optoelectronic application.

### 4 Conclusion

In summary, using ab initio simulation, the optoelectronic study of $(MA)_2NaBiX_6$ (MA= methylammonium; X=Cl,Br, I) hybrid double perovskites were investigated. The results depict tunable bandgap for all the investigated compound. This was found to be dependent on the solely dependent on the halogen content of the individual hybrid double perovskite material and as such make them suitable for different optoelectronic applications such as LCDs among others. More importantly, the I-based material reveal unique optoelectronic properties, including high absorption coefficient and high refractive index compared with the other compounds in the study as shown by the findings. These results predict the hybrid organic-inorganic double perovskite $(MA)_2NaBiI_6$ to be a good Pb-free material for different optoelectronic applications.

**Acknowledgement**
The authors wish to acknowledge the Cloud Computing Interface of Materials Square and the Computational Facility of the Centre for Food Technology and Research (CEFTER), the World Bank Africa Centre of Excellence in Benue State University, Makurdi, Nigeria.

**Author Contributions**
All authors have contributed equally.

**Funding**
The authors appreciate the doctoral fellowship of the Centre for Food Technology and Research (CEFTER), the World Bank Africa Centre of Excellence in Benue State University, Makurdi, Nigeria.

**Data Availability**
The authors confirm the data of this study are available within the article.

**Declaration**

**Conflict of interest**
The authors declare that they have no competing interests.

**Ethics approval**
Not applicable for that section as the study does not includes human subjects, human data or tissue, or animals.

**Informed consent**
All authors confirm their participation in this paper.

**Consent for publication**
All authors accept the publication rules applied by the journal.